\documentclass[twocolumn,preprintnumbers,superscriptaddress]{revtex4}
\usepackage{amsmath}
\usepackage{amssymb}
\usepackage{mathrsfs}
\usepackage{graphicx}
\usepackage{bm}
\usepackage[colorlinks=true,citecolor=blue,linkcolor=blue,urlcolor=blue]{hyperref}

\begin{document}
\title{Terahertz magneto-optical properties of graphene hydrodynamic electron liquid}

\author{L. F. Man}
\affiliation{School of Physics and Astronomy and Yunnan Key Lab for Quantum
Information, Yunnan University, Kunming 650091, China}

\author{W. Xu}\email{wenxu$_$issp@aliyun.com}
\affiliation{Micro Optical Instruments Inc., 518118 Shenzhen, China}
\affiliation{School of Physics and Astronomy and Yunnan Key Lab for Quantum
Information, Yunnan University, Kunming 650091, China}
\affiliation{Key Laboratory of Materials Physics, Institute of Solid State
Physics, Chinese Academy of Sciences, Hefei 230031, China}

\author{Y. M. Xiao}\email{yiming.xiao@ynu.edu.cn}
\affiliation{School of Physics and Astronomy and Yunnan Key Lab for Quantum
Information, Yunnan University, Kunming 650091, China}

\author{H. Wen}
\affiliation{Key Laboratory of Materials Physics, Institute of Solid State
Physics, Chinese Academy of Sciences, Hefei 230031, China}

\author{L. Ding}
\affiliation{School of Physics and Astronomy and Yunnan Key Lab for Quantum
Information, Yunnan University, Kunming 650091, China}

\author{B. Van Duppen}\email{ben.vanduppen@uantwerpen.be}
\affiliation{Department of Physics, University of Antwerp, Groenenborgerlaan 171,
B-2020 Antwerpen, Belgium}

\author{F. M. Peeters}
\affiliation{School of Physics and Astronomy and Yunnan Key Lab for Quantum
Information, Yunnan University, Kunming 650091, China}
\affiliation{Department of Physics, University of Antwerp, Groenenborgerlaan 171,
B-2020 Antwerpen, Belgium}

\date{\today}

\begin{abstract}
The discovery of the hydrodynamic electron liquid (HEL) in graphene
[D. Bandurin \emph{et al.}, Science {\bf 351}, 1055 (2016) and J. Crossno \emph{et al.},
Science {\bf 351}, 1058 (2016)] has marked the birth of the solid-state HEL
which can be probed near room temperature in a table-top setup. Here we
examine the terahertz (THz) magneto-optical (MO) properties of a graphene
HEL. Considering the case where the magnetic length $l_B=\sqrt{\hbar/eB}$
is comparable to the mean-free path $l_{ee}$ for electron-electron interaction
in graphene, the MO conductivities are obtained by taking a momentum balance
equation approach on the basis of the Boltzmann equation. We find that when
$l_B\sim l_{ee}$, the viscous effect in a HEL can weaken significantly the
THz MO effects such as cyclotron resonance and Faraday rotation. The
upper hybrid and cyclotron resonance magnetoplasmon modes $\omega_\pm$
are also obtained through the RPA dielectric function. The magnetoplasmons
of graphene HEL at large wave-vector regime are affected by the
viscous effect, and results in red-shifts of the magnetoplasmon frequencies.
We predict that the viscosity in graphene HEL can affect strongly the magneto-optical
and magnetoplasmonic properties, which can be verified experimentally.
\end{abstract}

\maketitle
\section{Introduction}
Hydrodynamic electron liquid (HEL) is a classic and important phenomenon in physics.
The viscous effect between different fluid layers in gas or liquid had been studied
even before the birth of quantum mechanics \cite{Landau66}. It was found that akin
to liquid and gas, the electrons in a Fermi-liquid can also exhibit features of
hydrodynamics \cite{Baym91}, where the effect of viscosity is induced by many-body
interactions in different electronic layers. The viscous effect in an electron gas
system contributes an extra resistance in additional to scattering from impurities and
phonons. Previously, HELs were observed mainly in the realm of high-energy physics
such as quark-gluon plasma at very high temperatures \cite{Jacak12} and atomic Fermi
gasses at very low temperatures \cite{Elliott14}. Since its discovery, graphene has attracted lots of attention
due to its unique and novel physical properties \cite{Novoselov04}. In particular, in 2016 two research teams
demonstrated experimentally that the solid-state HEL can be realized in graphene under
certain conditions \cite{Bandurin16,Crossno16} and that it is an
excellent material to investigate hydrodynamic flow of electrons.

Such important and exciting findings have marked the birth of a new subfield of
solid-state HEL from which one can explore the new physics of HEL near room temperature
in a table-top setup. Up to now, the solid-state HELs have been probed experimentally
by using mainly transport measurements \cite{Bandurin16,Crossno16,Lucas18,Ku20}.
One of the major advantages for a solid-state HEL is that we can employ optical
techniques, which are contact-less and nondestructive, to study and characterize the
hydrodynamic properties of a HEL. For an electron liquid to show hydrodynamic behavior,
it is crucial that the electron-electron (e-e) interactions are significant \cite{Torre15},
which means that the mean-free path induced by electron-electron interaction, $l_{ee}$,
should satisfy $l_{\mathrm{ee}}<l_\mathrm{s}$,  $L$, and
$l_{\mathrm{ee}}< v_\mathrm{F}/\omega$ \cite{Principi16}, for transport and optical
measurements, respectively. Here, $L$ is the dimension of the sample, $l_s$ corresponds
to the mean-free path caused by electronic scattering centers such as phonons and
impurities, $v_\mathrm{F}=10^6$ m/s is the Fermi velocity of graphene, and $\omega$
is the radiation photon frequency. In a graphene-based HEL, $l_{\mathrm{ee}}$ can be smaller than
0.1 $\mu$m \cite{Bandurin16}, $L$ can be easily of the order of millimeters and $l_s$
can be of the order of $\mu$m \cite{Mayorov11}. Therefore, it is no surprise that graphene
can be taken as the platform to probe the HEL via transport
measurements \cite{Bandurin16,Crossno16}. It should be noticed that the above relation
limits the range of validity for hydrodynamic description of a HEL to the low-frequency
range. Thanks to the availability of terahertz (THz) technology \cite{Lee08}, one can
employ the THz technique such as THz time-domain spectroscopy (TDS) \cite{Lee08} to probe and
study the viscosity and its frequency dependence by means of optical experiments.
When $\omega\sim 1$ THz, $v_\mathrm{F}/\omega\sim\mu$m so that the condition required for
the optical measurement of graphene-based HEL can be satisfied. From a physics viewpoint,
to probe the viscosity in a HEL we rely on the excitation and absorption of plasmons
in graphene, which is a direct consequence of e-e interaction. It is known that in
the absence of a magnetic field $B$, the optical conductivity for an electron gas
in the presence of viscosity can be written simply as \cite{Principi16}
\begin{equation}\label{opcon}
\sigma(q,\omega)={\sigma_0\over 1-i\omega\tau+\nu q^2\tau},
\end{equation}
where $\omega_\nu=\nu q^2$ is the characteristic frequency for a HEL and we call it the
viscous frequency, $\sigma_0=e^2 n_e v_\mathrm{F} \tau/(\hbar k_\mathrm{F})$ is
the dc conductivity of graphene at $B=0$, $n_e$ is the electron density, $\tau$
is the electronic momentum relaxation time, $\nu$ is the viscosity, $q$ is the
wave vector of elementary electronic excitation induced via e-e interaction and
$k_\mathrm{F}=\sqrt{\pi n_e}$ is the Fermi wave vector for graphene. From Eq. \eqref{opcon}, we learn that only excitations
with a non-zero momentum $q$ are sensitive to viscosity-induced relaxation
in optical measurement.

Recently, the study of the optoelectronic properties of graphene in the presence
of a magnetic field was presented such as the Landua levels, cyclotron
resonance, Faraday rotation, ellipticity, and
magnetoplasmons \cite{Jiang07,Tymchenko13,Crassee12,Yan12}.
In the presence of a static magnetic field applied perpendicularly to the
graphene flake, plasmons and cyclotron excitations would hybridize which can
lead to the formation of magnetoplasmons. In 2D electron gas, the magneto plasmon
has a dispersion relation $\omega^2=\omega^2_{c}+\omega^2_{p}$ which is called the
upper hybrid mode \cite{Ando82,Roldan09} where $\omega_c$ and $\omega_{p}$ are
cyclotron frequency and plasmon frequency, respectively. It has been shown that
the viscosity of graphene-based electron liquid is of the order of $\nu\approx$ 0.1
m$^2$s$^{-1}$ \cite{Bandurin16}. In the presence of magnetic field, the off-diagonal
part of viscous response coefficient is called Hall viscosity. The Hall viscosity
in graphene has been measured \cite{Berdyugin19} and it is smaller than the kinetic viscosity
at relatively weak magnetic field \cite{Narozhny19}. By the way, the Hall viscosity is also
the coefficient of $q^2$ in the small-$q$ limit \cite{Sherafati19}. Thus, in this
study we only consider the effect of the kinetic viscosity.
From Eq. \eqref{opcon}, we see that the viscosity in graphene HEL would affect
the optical conductivity at different
finite wavevector $q$. We predict that the viscosity can also affect the
magneto-optical properties of graphene HEL. The viscosity effect
in the presence of magnetic field can be probed via THz magneto-optical measurements \cite{Mei18,wxu2}.

In this work, we examine the magneto-optical (MO) properties of a
graphene-based HEL. Our approach is developed on the basis of a semi-classic
Boltzmann equation and random phase approximation (RPA). We will present and discuss the
results obtained for the magneto-optical conductivity, Faraday rotation angle and ellipticity, and the
magneto-plamson modes of a graphene HEL.

\section{Theoretical approach}
In this study, we consider the case of a relatively weak magnetic field which does
not induce Landau quantization and the corresponding magnetic length
$l_B=\sqrt{\hbar/eB}$ is comparable to $l_{ee}$ in graphene. We employ the semi-classic Boltzmann
equation (BE) approach to evaluate the MO conductivity of a graphene-based HEL in
the presence of the viscosity effect. Recently, the electronic hydrodynamics in graphene has been
investigated in detail on the basis of the kinetic theory to obtain the generalized Navier-Stokes
equation and the explicit expressions for the shear and Hall viscosity \cite{Narozhny19Rev}.
In this theoretical work \cite{Narozhny19Rev}, a two-band model has been used for arbitrary doping levels
of graphene where the infinite number of particles in the filled band were considered.
In the present study, we consider the case where the conducting carriers are only electrons and take
the viscous effect as an input parameter to show how it would affect the magneto-optical properties
of a graphene hydrodynamics system. Here, high quality samples are considered and the
electrons are distributed uniformly in the graphene film. Moreover, the temperature is
uniform and the chemical potential can be tuned via, e.g., applying a gate voltage.
In the presence of a weak radiation field, we assume the spatial inhomogeneities of
the distribution can be safely
ignored due to fast relaxation processes via, e.g., the electron-electron (e-e) scattering
which occurs in homogenous scattering centers. For simplification of our model, the
distribution function therefore would not depend on spatial coordinate since the
temperature and the chemical potential for the system are with no gradients.
The time-dependent semiclassic BE \cite{Xu97} for the homogeneous electrons in
conduction band of graphene is given as
\begin{align}\label{bolz}
{1\over \hbar}&[{\bf F}_1(t)+{\bf F}_2(t)]
\nabla_{\bf k} f_e({\bf k},t)+{\partial f_e({\bf k}, t)\over \partial t}\nonumber\\
&=g_sg_v\sum_{{\bf
k}'} [F({\bf k}',{\bf k};t)-F({\bf k},{\bf k}';t)],
\end{align}
where ${\bf k}=(k_x,k_y)$ is the electron wave vector, $f_e({\bf k},t)$ is the momentum
distribution function (MDF) for an electron in a state $|{\bf k}\rangle$ and at a time $t$,
$g_s=2$ and ${\it g_v}=2$ count respectively for spin and valley degeneracies, and
$F({\bf k},{\bf k'};t)=f_e({\bf k},t)W({\bf k},{\bf k'})$ is the collision term induced
by electronic scattering centers with $W({\bf k},{\bf k'})$ being the electronic
transition rate. Moreover, the force term here includes two contributions from,
respectively, the applied external field ${\bf F}_1(t)$ and the frictional
force \cite{Kubo92}: ${\bf F}_2(t)=-\gamma q^2 {\bf v}(t)$ with ${\bf v}(t)=[v_x(t),v_y(t)]$
being the electron drift velocity which also depends little on spatial coordinate in
a homogeneous graphene system, with $q$ being the wave vector of the elementary electronic
excitation via e-e interaction and $\gamma$ the viscosity coefficient. We consider
that the HEL is subjected simultaneously to a linearly polarized radiation field
and a static magnetic field $B$ in the Faraday geometry. Namely, the magnetic and radiation
fields are applied perpendicularly to the 2D plane (taken as the $xy$-plane) of the
graphene flake and the polarization of the light field is taken along the $x$-direction.
In this geometry, due to the coupling of the magnetic and radiation fields, the
effects of cyclotron resonance and the rotation of the light polarization by the
HEL can be observed. We can write the electric field component of the radiation
field as ${\bf E}(t)=F_0(1,0,0)e^{-i\omega t}$ and the magnetic field as ${\bf B}=(0,0,B)$
and, thus, ${\bf F}_1(t)=-e{\bf E} (t)-e{\bf v}(t)\times {\bf B}$. Equation \eqref{bolz} along
the $[x, y]$ direction then becomes
\begin{align}\label{bolz2}
-{1\over\hbar}\Big[&A_x(t){\partial f_e({\bf k},t)\over
\partial k_x},A_y(t){\partial f_e({\bf k},t)\over  \partial k_y}\Big]
+{\partial f_e({\bf k}, t)\over \partial t}\nonumber\\
&=g_sg_v\sum_{{\bf
k}'} [F({\bf k}',{\bf k};t)-F({\bf k},{\bf k}';t)],
\end{align}
with $A_x(t)=eF_0e^{-i\omega t}+eBv_y(t)+\gamma q^2 v_x(t)$ and
$A_y(t)=-eBv_x (t)+\gamma q^2 v_y(t)$.
For the first moment, the momentum-balance equation (MBE) can be derived by
multiplying $g_sg_v\sum_{\bf k}k_\alpha $ to both sides of the BE given
by Eq. \eqref{bolz2}, where $\alpha=(x,y)$. It should be noted that the main effect
of ${\bf F}_1(t)$ is to cause {\it a} drift velocity ${\bf v}(t)$ of the
electron in a HEL. As a result, the electron wave vector in the MDF is shifted
by \cite{Xu05,Xu91,Xu09} ${\bf k}\to {\bf k}^*(t)={\bf k}-(k_\mathrm{F}/v_\mathrm{F}){\bf v}(t)$.
Thus, the MBE derived from the BE becomes
\begin{equation}
 {n_ek_\mathrm{F}\over v_\mathrm{F}}\dot{v}_\alpha (t)+{n_e\over \hbar}A_\alpha(t)
 =16 \sum_{{\bf k}',{\bf k}}(k_\alpha'-k_\alpha)
f_e[{\bf k}^*(t)]W({\bf k},{\bf k}'),
\end{equation}
where $n_e=g_sg_v\sum_{\bf k}f_e({\bf k})$ and the dependence of the MDF on time
is mainly through the velocity ${\bf v}(t)$. For the case of a
relatively weak radiation field $E(t)$, the drift velocity of the
electron is relatively small so that we can make use of the expansion
\begin{equation}
f_e[{\bf k}^*(t)]\simeq f_e({\bf k})-{k_\mathrm{F}\over v_\mathrm{F}}
\big[v_x(t){\partial f_e({\bf k}) \over
\partial k_x}, v_y(t){\partial f_e({\bf k}) \over
\partial k_y}\big].
\end{equation}
Due to the symmetry of the electronic energy
spectrum of graphene, i.e., $E({\bf k})=\hbar v_\mathrm{F} k$, we have $\sum_{{\bf k}',{\bf
k}}(k_\alpha'-k_\alpha)f_e({\bf k})W({\bf k},{\bf k}')=0$ and, therefore, we have
\begin{equation}
{\partial v_\alpha(t)\over \partial t}+{v_\mathrm{F}A_\alpha(t)\over
\hbar k_\mathrm{F}}=-{v_\alpha(t)\over \tau},
\end{equation}
where the momentum relaxation time is determined by
\begin{equation}\label{mom}
{1\over \tau}={16\over n_e}\sum_{{\bf k}',{\bf k
}}(k_\alpha'-k_\alpha)W({\bf k},{\bf k}'){\partial f_e({\bf k})
\over\partial k_\alpha},
\end{equation}
in which the temperature dependence is included in the electron
distribution function and in the electronic scattering mechanisms.
The momentum relaxation time is contributed from electronic scattering
centers such as impurities, phonons, surface roughness, etc.
In our previous work \cite{Dong08,Xu09}, by using the momentum-balance
equation approach on the basis of the BE at $B=0$ we had discussed the
temperature dependence of the carrier density and the transport lifetime
(or momentum relaxation time) in graphene. The obtained theoretical
results reproduced the experimental results
measured via transport experiment \cite{Bandurin16} since
the DC conductivity is given by $\sigma_0=e^2 n_e v_\mathrm{F} \tau/(\hbar k_\mathrm{F})$. Furthermore, if we take the MDF $f_e({\bf k})$ in Eq. \eqref{mom} to be approximately as the statistical electron energy distribution function such as the Fermi-Diraction through: $f_e({\bf k})\simeq f_e[E({\bf k})]$ with $f_e (x)$ being the Fermi-Dirac function, the dependence of $\tau$ upon the temperature and chemical potential can be further included.
For weak radiation field $E(t)=F_0e^{-i\omega t}$, the electrons in graphene
respond linearly to the radiation so that $v_\alpha (t)=v_\alpha e^{-i\omega t}$.
Thus, we can get the steady-state electron drift velocity ${\bf v}=(v_x, v_y)$ and
the current density ${\bf J}=-n_e e {\bf v}$. After applying the current-voltage
matrix ${\bf J}=\sigma {\bf E}$ with $\sigma$ being the
conductivity tensor and ${\bf E}=F_0(1,0,0)$, the
longitudinal and transverse MO conductivities of HEL are obtained as, respectively,
\begin{equation}
\sigma_{xx}(q,\omega)=\sigma_0{1+\omega_v\tau-i\omega\tau \over
(1+\omega_\nu\tau-i\omega\tau)^2+(\omega_c\tau)^2},
\end{equation}
\begin{equation}
\sigma_{xy}(q,\omega)=-\sigma_0{\omega_c\tau \over
(1+\omega_\nu\tau-i\omega\tau)^2+(\omega_c\tau)^2},
\end{equation}
where $\omega_c=eBv_\mathrm{F}/(\hbar k_\mathrm{F})$ is
the cyclotron frequency of graphene and
$\omega_\nu=\gamma q^2 v_\mathrm{F}/(\hbar k_\mathrm{F})$. If we write
$\omega_\nu=\nu q^2$, then effective viscosity for a graphene HEL is
$\nu=\gamma k_\mathrm{F}/(\hbar v_\mathrm{F})\sim n_e^{1/2}$. When $B=0$,
$\sigma_{xy}(\omega)=0$ and $\sigma_{xx}(\omega)= \sigma(\omega)$, as given by Eq. \eqref{opcon}.

From $\sigma_{xx}(q,\omega)$ and $\sigma_{xy}(q,\omega)$, we can obtain the
right-handed circular (RHC) and left-handed circular (LHC) MO conductivities
via \cite{Palik70}
\begin{equation}
\sigma_\pm (q,\omega)=\sigma_{xx}\pm i \sigma_{xy}={\sigma_0\over
1+\omega_\nu\tau-i(\omega\mp\omega_c)\tau}.
\end{equation}

Now we look at the experimental aspects of the measurement by using THz TDS technique.
In a conventional THz TDS setup \cite{Mei18} the THz radiation is normally linearly
polarized. One often applies the linear polarizer to measure the transmission coefficients
at polarization angles $\pm 45^o$ to the polarization direction of the incident THz
beam \cite{Mei18}. If the electric field component of the THz beam transmitted through
the sample (graphene/substrate) or the substrate is $E_{\pm45^o,j}(\omega)$ with $j$
representing the result measured for the sample or the substrate, then the corresponding
right-hand circularly (RHC) and left-hand circularly (LHC) polarized light fields can be obtained
according to Fresnel theory \cite{Morikawa06},
\begin{equation}
\left[
\begin{array}{cc}
E_{+,j}(\omega) \\
E_{-,j}(\omega)
\end{array}\right]={1\over 2}
\left(
\begin{array}{cc}
i-1 & i+1 \\
i+1 & i-1
\end{array}
\right)\times
\left[
\begin{array}{cc}
E_{+45^\mathrm{o},j}(\omega) \\
E_{-45^\mathrm{o},j}(\omega)
\end{array}\right].
\end{equation}
Thus, the transmission coefficients of the RHC and LHC components for graphene
HEL can be determined by
\begin{equation}
t_\pm (\omega)=E_{\pm,\mathrm{sample}}(\omega)/E_{\pm,\mathrm{substrate}}(\omega).
\end{equation}

The relationship between the transmission coefficients of the RHC and LHC
components and the RHC and LHC conductivities is \cite{Chiu76}
\begin{equation}
t_\pm (\omega)={2n_0\over n_0+n_s+Z_0\sigma_\pm (q,\omega)}=|t_\pm(\omega)|e^{i\zeta_\pm (\omega)},
\end{equation}
where $n_0=3.6$ and $n_s=2.2$ are the indices of refraction of the substrate and graphene
respectively, $Z_{0} \approx 377 \ \Omega$ is the impedance of free space, and
$|t_\pm(\omega)|$ and $\zeta_\pm (\omega)$ are the module and the phase angle
of the transmission coefficient, respectively. It should be noted
that the presence of the dielectric substrate may influence the hydrodynamic
regime in graphene. It was shown that graphene on hexagonal boron nitride (h-BN) substrate
can have electronic mobility and carrier inhomogeneity that are almost an order of magnitude
better than devices on SiO$_2$ substrate \cite{Dean10,Mayorov11}. To minimize the disorder for sustaining
hydrodynamic regime, the graphene film is usually encapsulated between the h-BN crystals for experimental
settings \cite{Bandurin16,Berdyugin19,Crossno16}.

The Faraday rotation angle
$\theta(\omega)$ and the ellipticity $\eta(\omega)$ are related to the
transmission coefficients through \cite{OConnell82}
\begin{equation}
{t_-(\omega)\over t_+(\omega)}={1-\eta(\omega)\over 1+\eta(\omega)}e^{-2i\theta(\omega)}.
\end{equation}
Therefore, we obtain the Faraday rotation angle
\begin{align}
\theta(\omega)={1\over 2}[\zeta_+ (\omega)-\zeta_- (\omega)]={1\over 2}\tan^{-1}\frac{r(b_-d_+-d_-b_+)}{r^2 b_-b_++d_-d_+},
\end{align}
where $a=1+\omega_v\tau$, $b_{\pm}=(\omega\mp\omega_c)\tau$, $r=Z_0\sigma_0/(n_0+n_s)$, and $d_{\pm}=(a^2+b_\pm^2+ra)$.
Meanwhile, the ellipticity is given by
\begin{equation}
\eta(\omega)={|t_+(\omega)|-|t_-(\omega)|\over |t_+(\omega)|+|t_-(\omega)|}=\frac{c_+f_--c_-f_+}{c_-f_++c_+f_-},
\end{equation}
where $c_\pm=a^2+b_\pm^2$ and $f_\pm=\sqrt{d_\pm^2+r^2b_\pm^2}$.

Furthermore, the longitudinal conductivity has a direct relation to the density-density correlation
function $\sigma_{xx}(q,\omega)=ie^2\omega \chi_{nn}(q,\omega)/q^2$. The random-phase approximation (RPA) dielectric function can be written as \cite{Giuliani05}
\begin{equation}
\epsilon_{\mathrm{RPA}}(q,\omega)=1-v_q\chi_{nn}(q,\omega)=1+\frac{iq^2v_q}{e^2\omega}\sigma_{xx}(q,\omega),
\end{equation}
The magneto-plasmons can then be obtained by the real roots for the zeros of the real part of the dielectric
function $\mathrm{Re}\ \epsilon_{\mathrm{RPA}}(q,\omega)\rightarrow 0$. The imaginary part of $\epsilon_{\mathrm{RPA}}(q,\omega)$ relates directly the decay or damping of the electronic motion. Thus, we obtain two branches of the plasmon modes as
\begin{equation}
\omega^2_{\pm}=\omega^2_c+\frac{\omega_p^2}{2}-(\tau^{-1}+\nu q^2)^2\pm\sqrt{\frac{\omega_p^4}{4}-4\omega_c^2(\tau^{-1}+\nu q^2)^2},
\end{equation}
where $\omega_p=\sqrt{e^2E_\mathrm{F}q/(2\pi\epsilon_r\epsilon_0)}$ is the Dirac plasmon
frequency of graphene.

\section{Results and discussions}
\label{sec:results}

\begin{figure*}[t]
\includegraphics[width=13cm]{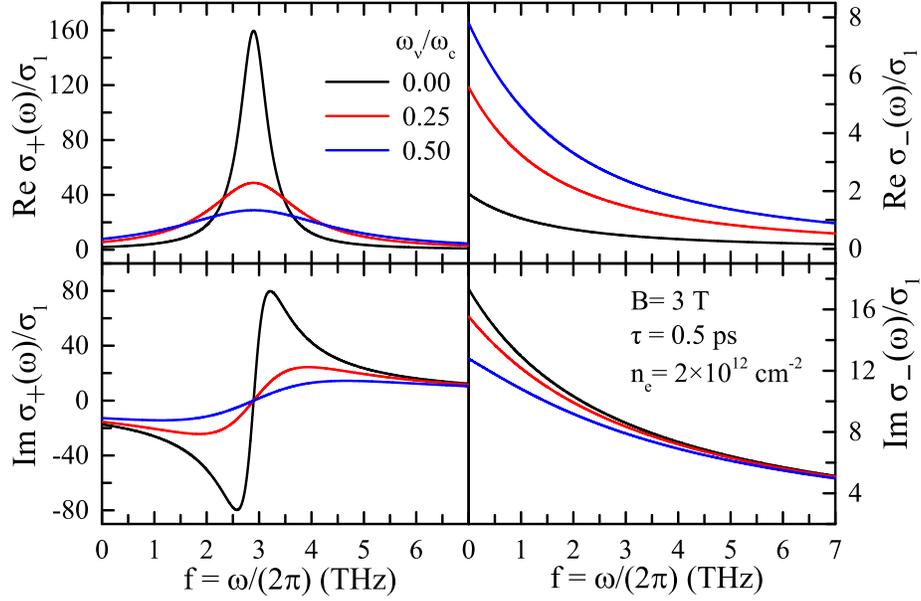}
\caption{Theoretical spectra of the real and imaginary parts of the RHC and
LHC MO conductivities, $\sigma_\pm (\omega)$, as a function of radiation frequency
$f=\omega/(2\pi)$ at the fixed magnetic field, electron density and electronic relaxation
time for different viscous frequencies. Here, the optical conductivity is
in an unit of $\sigma_1=e^2/4\hbar$.}\label{fig1}
\end{figure*}

\begin{figure*}[t]
\centering
\includegraphics[width=13cm]{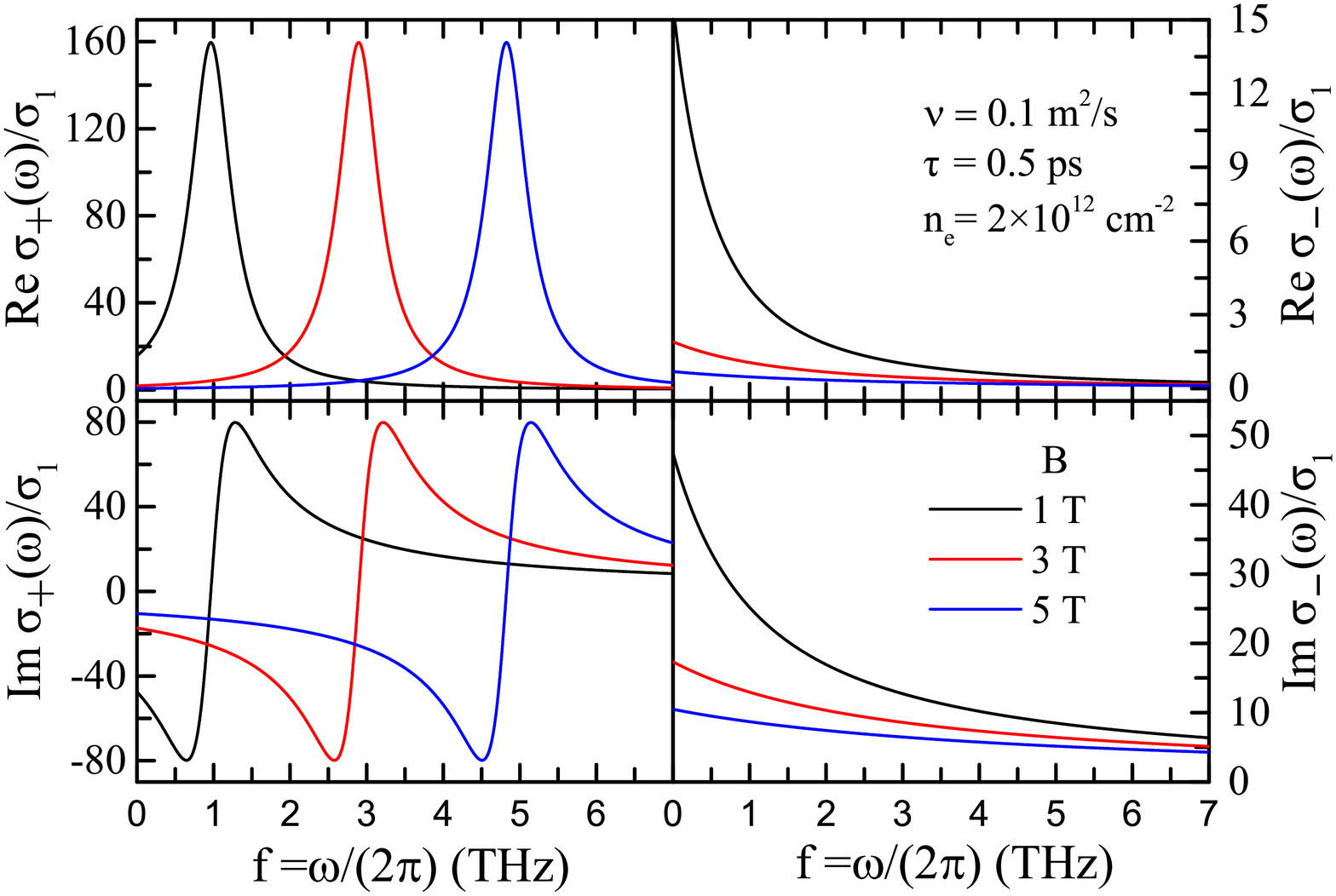}
\caption{The real and imaginary parts of the RHC and
LHC MO conductivities, $\sigma_\pm (\omega)$, as a function of radiation frequency
$f=\omega/(2\pi)$ at the fixed viscosity with wave vector $q=3\times10^5$ cm$^{-1}$, electron density and electronic relaxation time for different magnetic field strengths.}\label{fig2}
\end{figure*}

For numerical calculations, we use the typical carrier
density with the order $n_e\sim10^{12}$ cm$^{-2}$ for a ``higher-than-in-hone''
viscosity $\nu\approx$ 0.1 m$^2$s$^{-1}$ \cite{Bandurin16,Narozhny19} at a
temperature $T=220$ K. Ho and coworkers reported \cite{Ho18} that a large and robust hydrodynamic
window exists in graphene over a wide range of carrier density deviation from charge
neutrality and temperatures from 150 to 300 K, where the e-e interaction
is the dominant scattering mechanism and it can support both the plasmon
and single-carrier hydrodynamic regimes. In the presence of viscous effect in
a graphene HEL, the effect of viscosity on the optical conductivity and plasmon
should relative to a finite wave vector $q$. In terms of spectroscopic
measurement, such a momentum mismatch can be realized through, e.g., patterning the
gratings on graphene film \cite{Wenger16} which has been realized experimentally for the observation of the HEL effect in graphene \cite{Maier07,Zhu13,Jadidi15}.

\begin{figure}[t]
\centering
\includegraphics[width=8.6cm]{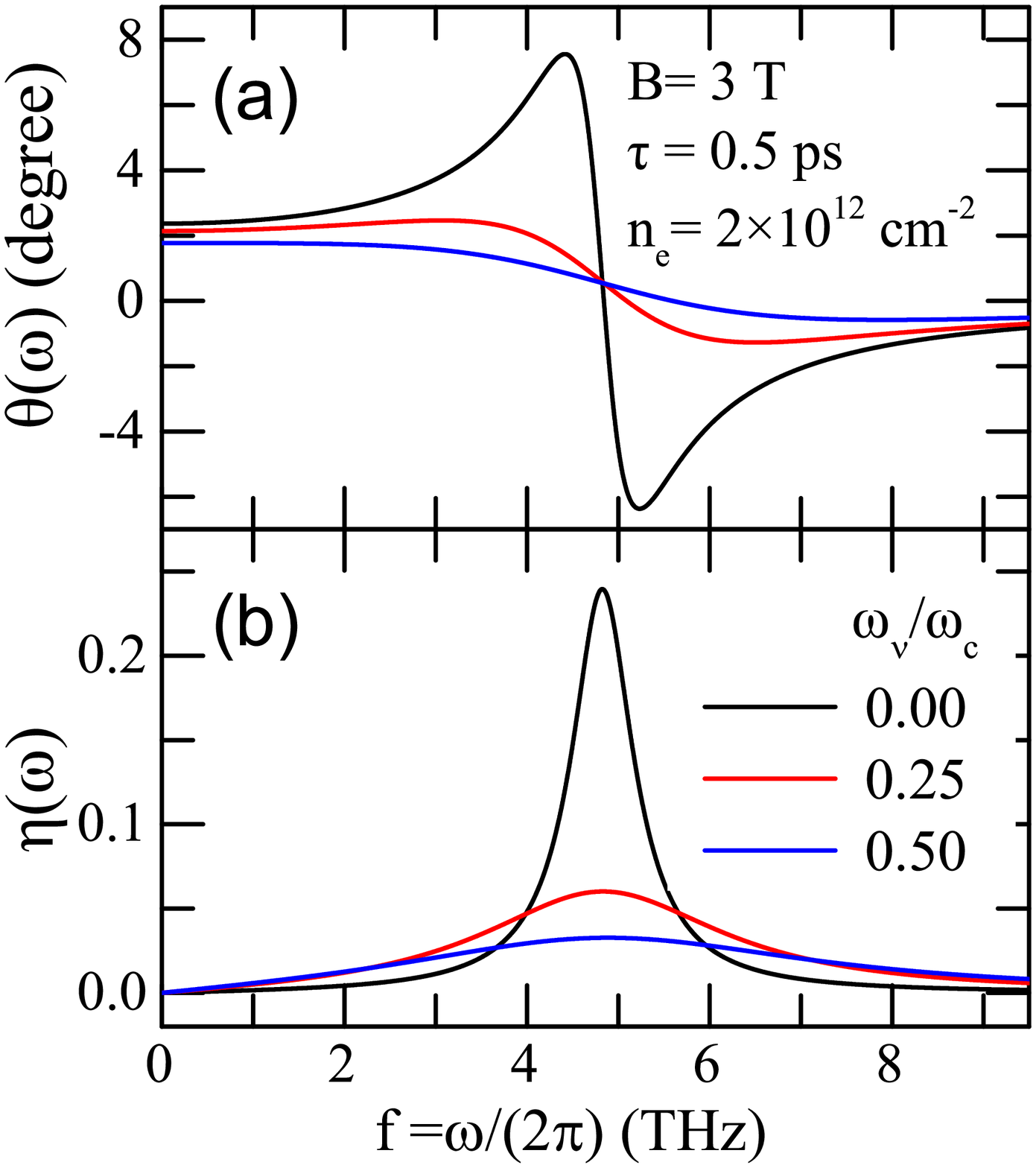}
\caption{Spectra of (a) the Faraday rotation angle $\theta (\omega)$ and (b)
the Faraday ellipticity $\eta (\omega)$ as a function of radiation frequency $f=\omega/(2\pi)$ at the fixed magnetic
field, electron density and electronic relaxation time for different viscous
frequencies.}\label{fig3}
\end{figure}

In conjunction with the MO THz TDS measurement, in Fig. \ref{fig1} we plot the
real and imaginary parts of the RHC and LHC conductivities, $\sigma_\pm(\omega)$,
as a function of radiation frequency $f=\omega/2\pi$ at the fixed magnetic field
$B=3$ T with a typical electron density $n_e=2\times 10^{12}$ cm$^{-2}$
for different viscous frequencies
$\omega_\nu$. For the calculation, we take a typical electronic relaxation time $\tau=0.5$ ps.
We note that, with the Drude-like MO conductivity as used in this study, the temperature
dependence is included in the momentum relaxation time given by Eq. \eqref{mom}
since $\tau$ is a functional form of the carrier density and temperature. The increase in
temperature can lead to a decrease in $\tau$ in graphene system due mainly to the electron-phonon scattering \cite{Bandurin16,Dong08}. The temperature dependence of $\tau$ can be determined through transport measurement and
the corresponding values can be included in the evaluation of the viscosity effect.
Similar to other electron gas systems \cite{OConnell82}, the cyclotron
resonance can be observed in the real and imaginary parts of $\sigma_+(\omega)$
when $\omega\sim\omega_c$. From Fig. \ref{fig1}, we find that the real and imaginary
parts of $\sigma_\pm (\omega)$ decrease with increasing viscous frequency
$\omega_\nu$ or the strength of the viscosity. This effect is akin to
the case of transport experiments for a HEL in which an extra resistance
is added to the total resistance in additional to that induced by electronic
scattering from impurities and phonons \cite{Pellegrino17}.
In Fig. \ref{fig2}, we show the real and imaginary parts of RHC and LHC
conductivities as a function of radiation frequency at the fixed viscous frequency,
electron density $n_e=2\times 10^{12}$ cm$^{-2}$ for different magnetic fields. The real and imaginary parts of RHC conductivities
blueshift with increasing magnetic field while the strength of LHC
conductivities decrease with increasing $B$.
In Fig. \ref{fig3} we present the corresponding Faraday ellipticity $\eta (\omega)$ and rotation
angle $\theta (\omega)$ as a function of radiation frequency at the fixed magnetic field,
electron density and electronic relaxation time for different viscous frequencies
$\omega_\nu$. We find, again, that both $\eta (\omega)$ and $\theta (\omega)$ decrease
with increasing $\omega_\nu$. Therefore, the viscosity can
weaken the magneto-optical absorption in a graphene HEL.

\begin{figure}[t]
\centering
\includegraphics[width=8.60cm]{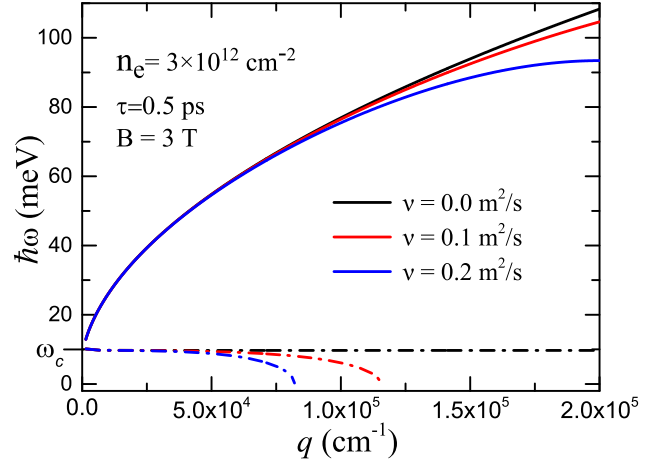}
\caption{The magneto-plasmon energy of a graphene HEL as a function of wave vector $q$
at the fixed electron density and magnetic field for different viscosities.
The solid and dash-dotted curves denote the upper hybrid mode $\omega_+$ and
the cyclotron resonance mode $\omega_-$, respectively.}\label{fig4}
\end{figure}

\begin{figure}[b]
\centering
\includegraphics[width=8.6cm]{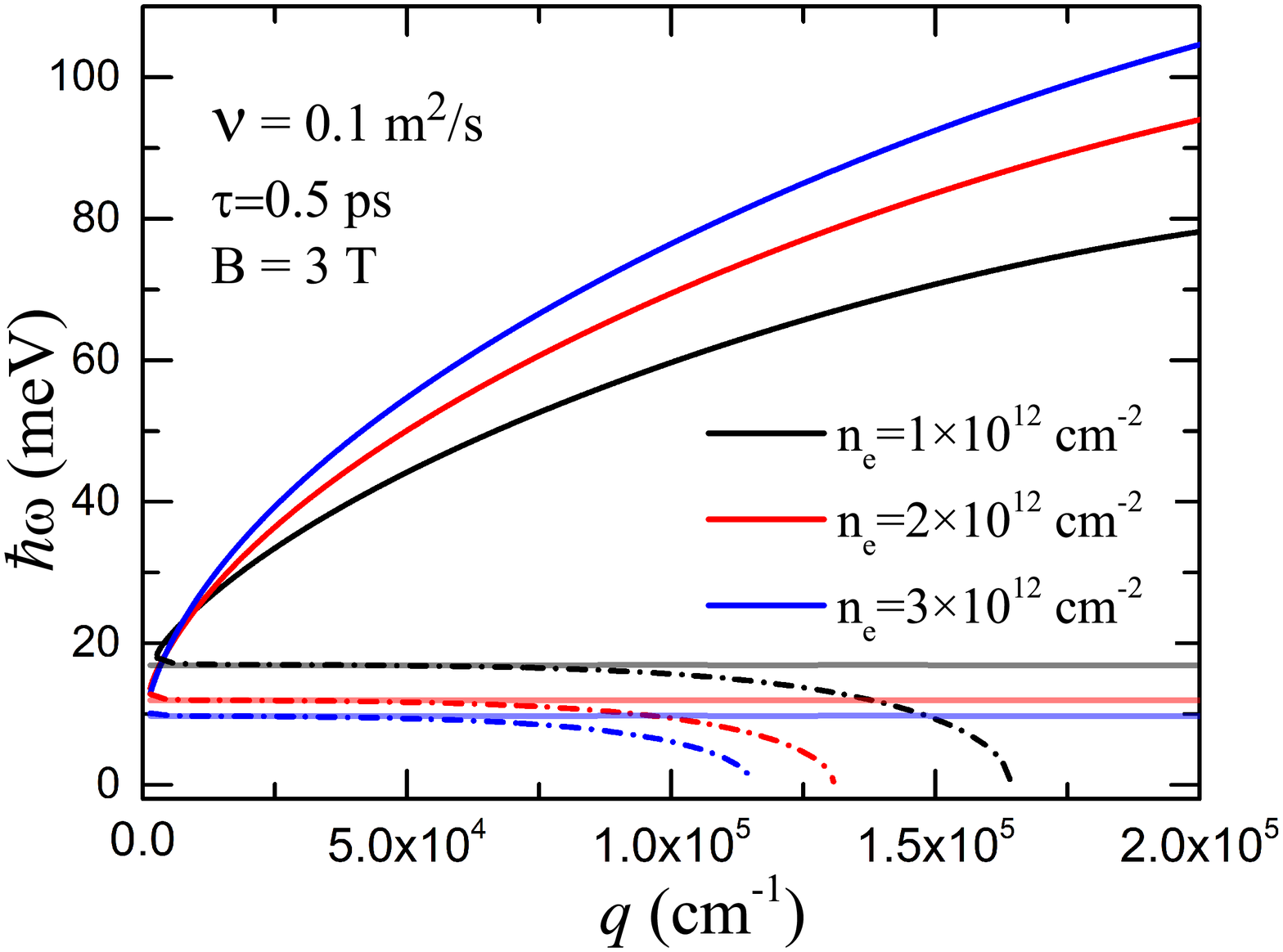}
\caption{The magnetoplasmon energy of a graphene HEL as a function of wave vector $q$
at the fixed viscosity and magnetic field for different electron densities.
The solid and dash-dotted curves denote the upper hybrid mode $\omega_+$ and
the cyclotron resonance mode $\omega_-$, respectively.  The translucence
solid horizonal lines are the cyclotron frequencies for the corresponding
cyclotron resonant modes.}\label{fig5}
\end{figure}

In Fig. \ref{fig4} we plot the magnetoplasmon energy as a function of wave vector $q$
at the fixed electron density, magnetic field strength for different viscosities.
Notice, there are two magnetoplasmon branches corresponding to the upper
hybrid mode $\omega_+$ and the cyclotron resonance mode $\omega_-$. In the
absence of viscosity $\nu=0$, the upper hybrid mode
$\omega_+\simeq\sqrt{\omega^2_c+\omega^2_p}$ \cite{Ando82} and the cyclotron resonance mode
$\omega_-\simeq\omega_c$ corresponds to the cyclotron resonance frequency.
In the collisionless limit (i.e., $\tau\rightarrow \infty$), the $\omega_-$ mode
vanishes in the small $q$ regime and only the $\omega_+$ mode exists \cite{Volkov16}.
The magnetoplasmon frequencies of the upper hybrid mode and cyclotron resonance mode
at large $q$ decrease with increasing viscosity. When the viscosity $\nu\neq0$,
the cyclotron resonance mode decreases with increasing $q$ and approaches to
zero at a cutoff wave vector. This mode with a frequency below $\omega_c$
attributes to the friction force which can slow down the electron movement owing to the
e-e interaction in the presence of the viscosity. The magnetoplasmon
dispersion at the fixed magnetic field strength,
viscosity for different electron densities is shown in Fig. \ref{fig5}.
The cyclotron frequency of graphene $\omega_c=eBv_\mathrm{F}/(\hbar k_\mathrm{F})$
depends on the Fermi wave vector $k_\mathrm{F}$. Thus, $\omega_c$ decreases with increasing
carrier density. With increasing the carrier density, the magnetoplasmon
frequency of the upper hybrid mode is higher and the cyclotron resonance
mode becomes lower in a larger wave vector $q$ regime.

\begin{figure}[t]
\centering
\includegraphics[width=8.6cm]{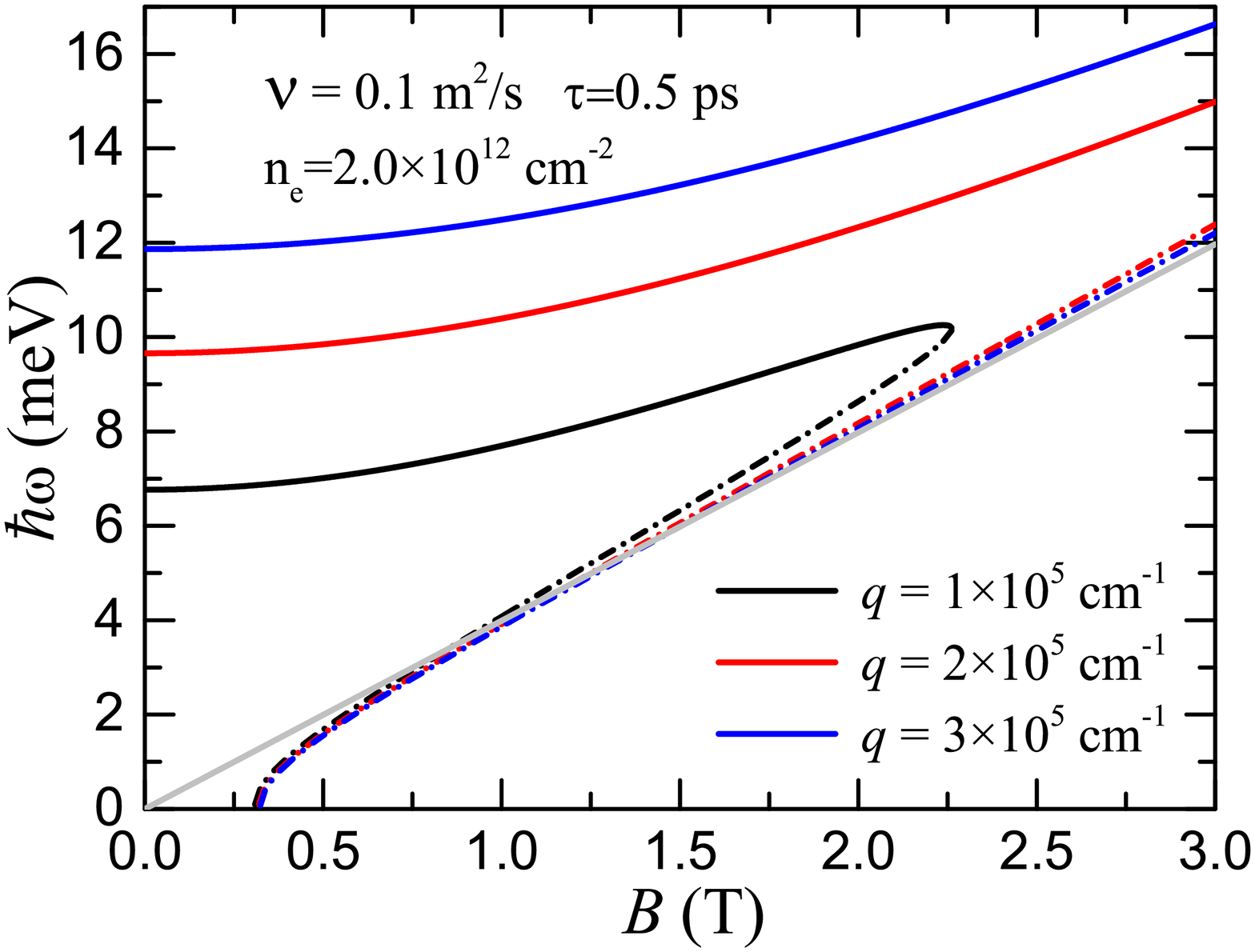}
\caption{The magnetoplasmon modes of a graphene HEL as a function of magnetic field $B$
at the fixed viscosity and electron density for different wave vectors $q$. The solid
and dashed-dotted curves denote the upper hybrid mode $\omega_+$ and
the cyclotron resonance mode $\omega_-$, respectively. Here, the gray line denotes
the cyclotron frequency $\hbar\omega_c$.}\label{fig6}
\end{figure}

In Fig. \ref{fig6} we plot the magnetoplasmon energy as a function of magnetic field
strength at the fixed carrier density and viscosity for different wave vectors $q$.
The two magnetoplasmon frequencies increase with increasing magnetic field
$B$. For a fixed wave vector $q=1\times10^5$ cm$^{-1}$, the upper hybrid mode and
the cyclotron resonance mode coincide at a cutoff $B$ point with increasing
the magnetic field $B$. We can see that the frequency of the cyclotron
resonance mode is slightly below the cyclotron frequency at small magnetic fields and higher than
the cyclotron frequency at large magnetic fields. We can clearly observe that the upper hybrid mode
show the $\omega_+\simeq\sqrt{\omega^2_c+\omega^2_p}$ dependence upon the magnetic
field strength.

In recent years, the THz TDS technique has been employed to investigate into the
optoelectronic properties of atomically thin electronic systems such as
graphene \cite{Mei18}, monolayer (ML) MoS$_2$ \cite{Wang19}, ML WS$_2$ \cite{Dong20},
ML hBN \cite{Bilal20}, etc. It had been demonstrated experimentally \cite{Mei18}
that through MO THz TDS measurement, we can obtained the spectra of the real and
imaginary parts of $\sigma_\pm(\omega)$ or $\sigma_{xx}(\omega)$ and
$\sigma_{xy}(\omega)$. The magnetoplasmon resonances could also be probed by
a variety of methods such as electron energy-loss spectroscopy, inelastic light
scattering, angle-resolved photoemission spectroscopy (ARPES), scanning
tunnelling spectroscopy \cite{Grigorenko12}, etc. From our theoretical calculation,
the viscousity in graphene-based HEL weakens the optical conductivity and
affects the magnetoplasmon properties at large wave vector $q$.

\section{Conclusions}
\label{sec:conclusions}

In this work, we consider a relatively weak magnetic field such that it
does not induce the Landau quantization and the magnetic length $l_B$
is comparable to $l_{ee}$ in graphene. In such a case, the cyclotron
resonance induced by intra-band electronic transition can be observed
in the MO conductivity and the Faraday rotation in THz
frequency regime. Because $l_B\sim l_{ee}$, the presence of viscosity
in a HEL can weaken significantly these characteristic MO effects.
We have investigated the magnetoplasmon modes in graphene HEL in which an
upper hybrid mode and a cyclotron resonance mode exist. Both
two modes are affected strongly by the viscous effect in the large wave vector
regime. The magnetoplasmon frequency of the cyclotron resonance mode
decreases and approaches to zero with decreasing wave vector.
From the results obtained for a relatively weak magnetic field,
one can expect that the magnetoplasmon effect occur at relatively
large wave vector regime and exhibit a frequency shift caused by the
viscosity, especially the $\omega_-$ mode which can be below
the cyclotron resonance frequency $\omega_c$.

In summary, the recent discovery of the hydrodynamic electron liquid
(HEL) in graphene has opened up a new subfield of solid-state HEL.
This allows us to explore the new physics of HEL near room temperature in
a table-top setup. In this work,
we have studied theoretically the terahertz (THz) magneto-optical (MO) and
magnetoplasmon properties of a graphene-based HEL. The present study has
been focused on the situation where the magnetic length $l_B=\sqrt{\hbar/eB}$
is comparable to the mean-free path $l_{ee}$ for electron-electron interaction
in graphene. We have obtained the longitudinal and transverse MO conductivities,
$\sigma_{xx}(\omega)$ and $\sigma_{xy}(\omega)$, by using a momentum balance
equation approach derived from a semiclassic Boltzmann equation in which the
frictional force induced by the viscosity is included. We also obtain
the right-handed circular and left-handed circular MO conductivities,
$\sigma_\pm (\omega)$, along with the Faraday rotation angle $\theta (\omega)$
and ellipticity $\eta (\omega)$. The magneto-plasmon modes have been evaluated using
the model of the RPA. The upper hybrid mode and the cyclotron
resonance mode are obtained and examined.

In conjunction to the MO THz TDS measurement, we have examined the dependence
of the spectra of the real and imaginary parts of $\sigma_\pm (\omega)$, along
with $\theta (\omega)$ and $\eta (\omega)$, on the viscous frequency. It has been found
that the viscous effect in a HEL can weaken significantly the THz MO effects
such as the cyclotron resonance and the Faraday rotation. The magnetoplasmon modes
in graphene HEL in the large wave vector regime are affected a lot by the
viscous effect, especially the red-shifts of the magnetoplasmon frequency can be observed.
We hope that the theoretical prediction in this study can be verified
experimentally and deepen our understanding of the magneto-optical properties
of graphene-based HEL.

\section*{ACKNOWLEDGMENTS}
This work was supported by the National Natural Science foundation
of China (U1930116, U1832153, U206720039, 12004331, 11847054), Shenzhen Science and Technology Program
(KQTD20190929173954826),
the PIFI program of the Chinese Academy of Sciences, and by Yunnan
Fundamental Research Projects (grand No. 2019FD134), BVD was supported
through a postdoc fellowship from the Research Foundation Flanders.

\end{document}